\documentclass[a4paper]{jpconf}
\usepackage{graphicx}
\usepackage{wrapfig,rotating}
\usepackage{epsfig}
\usepackage{cite}
\usepackage{color}
\begin{document}
\title{Data Preservation in High Energy Physics}

\author{David M. South, on behalf of the ICFA DPHEP Study Group}

\address{Deutsches Elektronen Synchrotron, Notkestra\ss e 85, 22607 Hamburg, Germany}

\ead{david.south@desy.de}

\begin{abstract}
Data from high--energy physics (HEP) experiments are collected with significant financial and human effort and
are in many cases unique. At the same time, HEP has no coherent strategy for data preservation and re--use, and many
important and complex data sets are simply lost. In a period of a few years, several important and unique experimental
programs will come to an end, including those at HERA, the b--factories and at the Tevatron.
An inter-experimental study group on HEP data preservation and long-term analysis
(DPHEP) was formed and a series of workshops were held to investigate this issue in a systematic way.
The physics case for data preservation and the preservation models established by the group are
presented, as well as a description of the transverse global projects and strategies already in place.

\end{abstract}

\section{Introduction}
\label{sec:intro}

\begin{wrapfigure}{l}{18.5pc}
\begin{center}
\vspace{-0.3cm}
  \includegraphics[width=18.5pc]{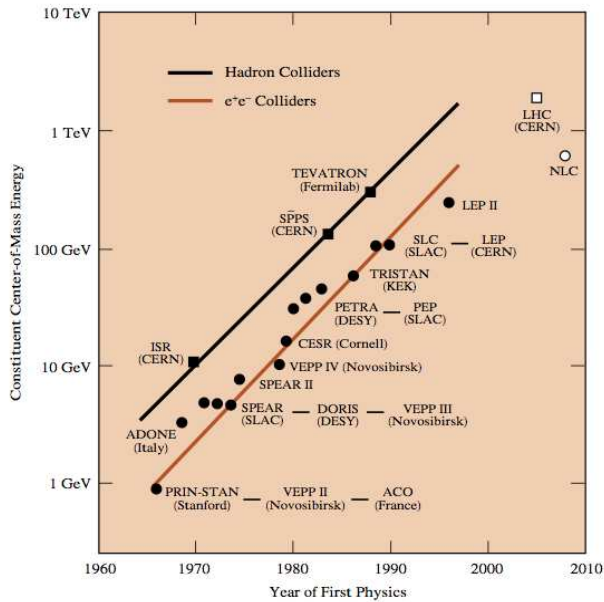}
\end{center}
\vspace{-0.3cm}
 \caption{\label{fig:history} A history of the constituent centre--of--mass energy of electron-positron and hadron 
   colliders, as a function of the year of first physics results~\cite{panofsky}.}
\vspace{-0.3cm}
\end{wrapfigure}

Since the $1950$s, physicists have constructed particle colliders to study the
building blocks of matter, where technological advances, as well as experimental
discoveries, have resulted in the construction of bigger and more powerful accelerators.
In most cases the next generation collider operates at a higher energy frontier or intensity
than the previous one.
This feature is displayed in figure~\ref{fig:history}, which shows the last $50$ years in
particle physics, where the clear trend to higher energies is visible in both
hadron--hadron and $e^{+}e^{-}$ colliders~\cite{panofsky}.


At the end of the first decade of the $21^{st}$ century, the focus is firmly
on the Large Hadron Collider (LHC) at CERN, which operates mainly as a 
$pp$ collider, currently at a centre--of--mass energy of $7$~TeV, where
the first significant physics results are now emerging~\cite{firstlhcresults}.
At the same time, a generation of other high energy physics (HEP) experiments
are concluding their data taking and winding up their physics programmes.
These include H1 and ZEUS at the world's only $e^{\pm}p$ collider
HERA (data taking ended July $2007$), BaBar at the $e^{+}e^{-}$ collider
at SLAC (ended April $2008$) and the Tevatron $p\bar{p}$ experiments
D{\O} and CDF, who are now due to stop data taking in  September $2011$~\cite{tevatron}.
The Belle experiment also recently concluded data taking at the KEK $e^{+}e^{-}$
collider, where upgrades are now ongoing until 2012~\cite{belle}.


The experimental data from these experiments still has much to tell us from
the ongoing analyses that remain to be completed, but it may also contain
things we do not yet know about.
The scientific value of long term analysis was examined in a recent
survey by the PARSE-Insight project~\cite{parse}, where around $70\%$ of over
a thousand HEP physicists regarded data preservation as very important or even crucial.
Moreover, the data from in particular the HERA and Tevatron experiments are
unique in terms of the initial state particles and are unlikely to be superseded
anytime soon, even considering such future projects as the LHeC~\cite{lhec}. 


It would therefore be prudent for such experiments to envisage some form of
conservation of their respective data sets.
However, HEP has little or no tradition or clear current model of long term
preservation of data in a meaningful and useful way.
It is likely that the majority of older HEP experiments have in fact simply
lost the data: misplaced, accidentally deleted, or if still existing only in some
unusable state.
The preservation of and supported long term access to the data is generally
not part of the planning, software design or budget of a HEP  experiment and
for the few known preserved HEP data examples, in general the exercise has
not been a planned initiative by the collaboration but a push
by knowledgeable people, usually at a later date.
The distribution of the data complicates the task, with potential headaches
arising from ageing hardware where the data themselves are stored, as well
as from unmaintained and outdated software, which tends to be under the
control of the (defunct) experiments rather than the associated HEP
computing centres.


To address this issue in a systematic way, a study group on Data
Preservation and Long Term Analysis in High Energy Physics, DPHEP, was
formed at the end of $2008$~\cite{dpheporg}.
The aims of the study group include to confront the data models,
clarify the concepts, set a common language, investigate the technical aspects,
and to compare with other fields such as astrophysics and those handling large data sets.
The experiments BaBar, Belle, BES-III, CLAS, CLEO, CDF, D{\O}, H1
and ZEUS and the associated computing centres at DESY (Germany), Fermilab (USA),
IHEP (China), JLAB (USA), KEK (Japan) and SLAC (USA) are all represented in DPHEP.

\begin{wrapfigure}{r}{17.0pc}
\begin{center}
\vspace{-0.6cm}
  \includegraphics[height=17.0pc, angle=90]{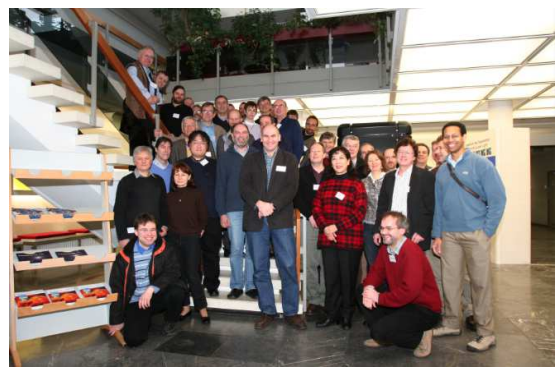}
\end{center}
\vspace{-0.3cm}
 \caption{\label{fig:people} Participants of the first DPHEP workshop at DESY, January 2009.}
\vspace{-0.3cm}
\end{wrapfigure}

A series of workshops~\cite{dphep1,dphep2,dphep3,dphep4}
have taken place over the last two years, beginning at DESY in January $2009$
and most recently at KEK in July $2010$.
The study group is officially endorsed with a mandate by the International
Committee for Future Accelerators, ICFA~\cite{icfa} and the first DPHEP recommendations
were published in $2009$, summarising the initial findings and setting out future working
directions~\cite{dpheppub1}.
The aims of the study group have also been presented to a wider physics audience via
seminars, conferences and publications in
periodicals~\cite{southmoriond09,bethkeqcd10,dphepcerncour}.
The role of the DPHEP study group is to provide international coordination
of data preservation efforts in high energy physics and to provide a set of
recommendations for past, present and future HEP experiments.


In the following, the physics case for data preservation is examined, followed by the
different models for data preservation identified by the study group.
Current inter-experimental data preservation initiatives are then presented, followed
by some words on governance and structures, before finally concluding with an outlook and
summary of future working directions.

\section{The Physics Case for Data Preservation}
\label{sec:physcase}

The motivation behind data preservation in HEP should have its roots in physics.
One of the main assumptions concerning experimental HEP data is that older data
will always be superseded by that from the next generation experiment, usually at the
next energy frontier.
However, this is not always the case as illustrated by the two following
recent, notable examples of analysis of older HEP data.

\begin{figure}[h]
\begin{minipage}{18.5pc}
\includegraphics[width=18.5pc]{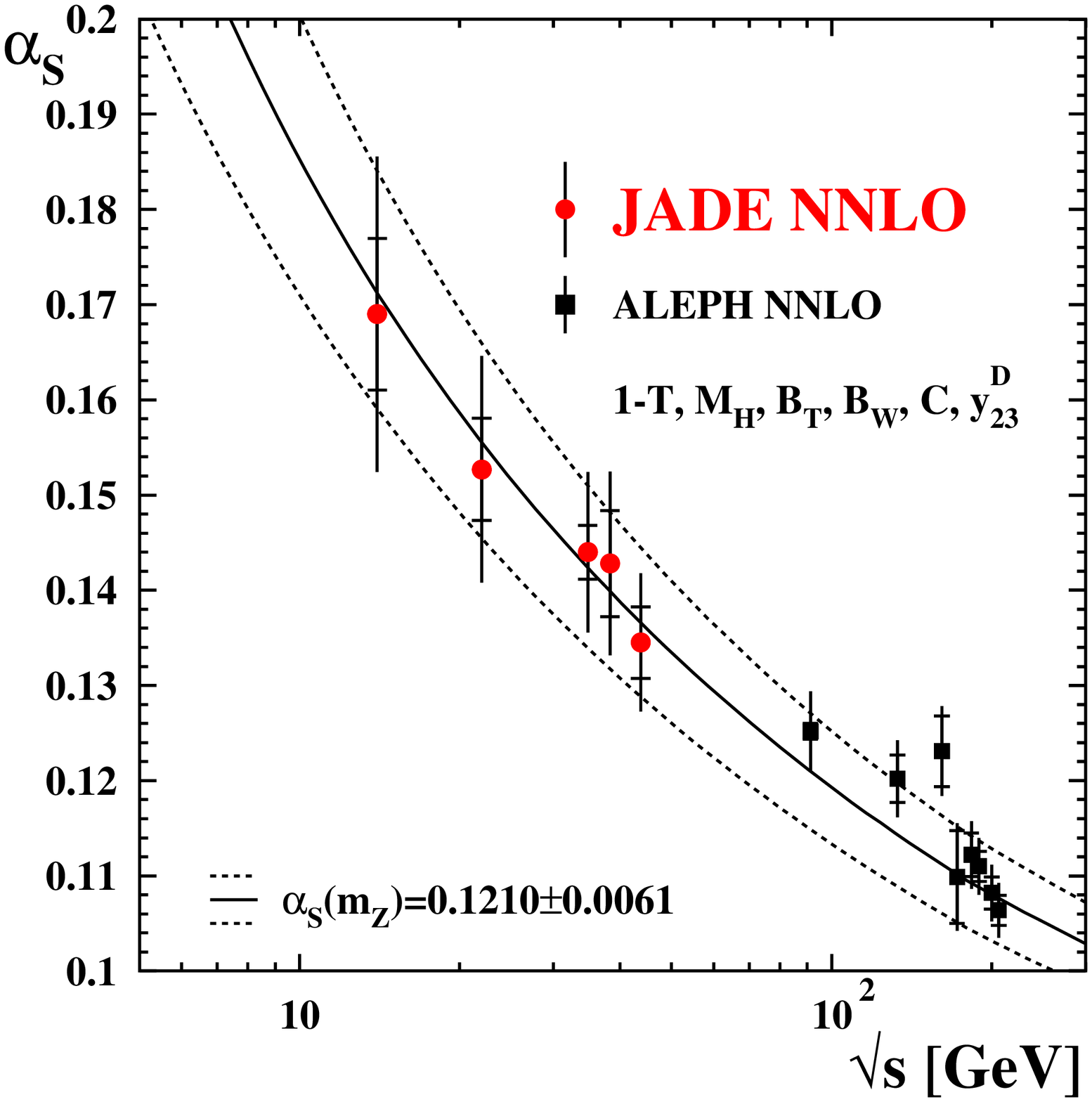}
\end{minipage}
\hspace{2pc}
\begin{minipage}{18.5pc}
\vspace{-0.75cm}
\includegraphics[width=18.5pc]{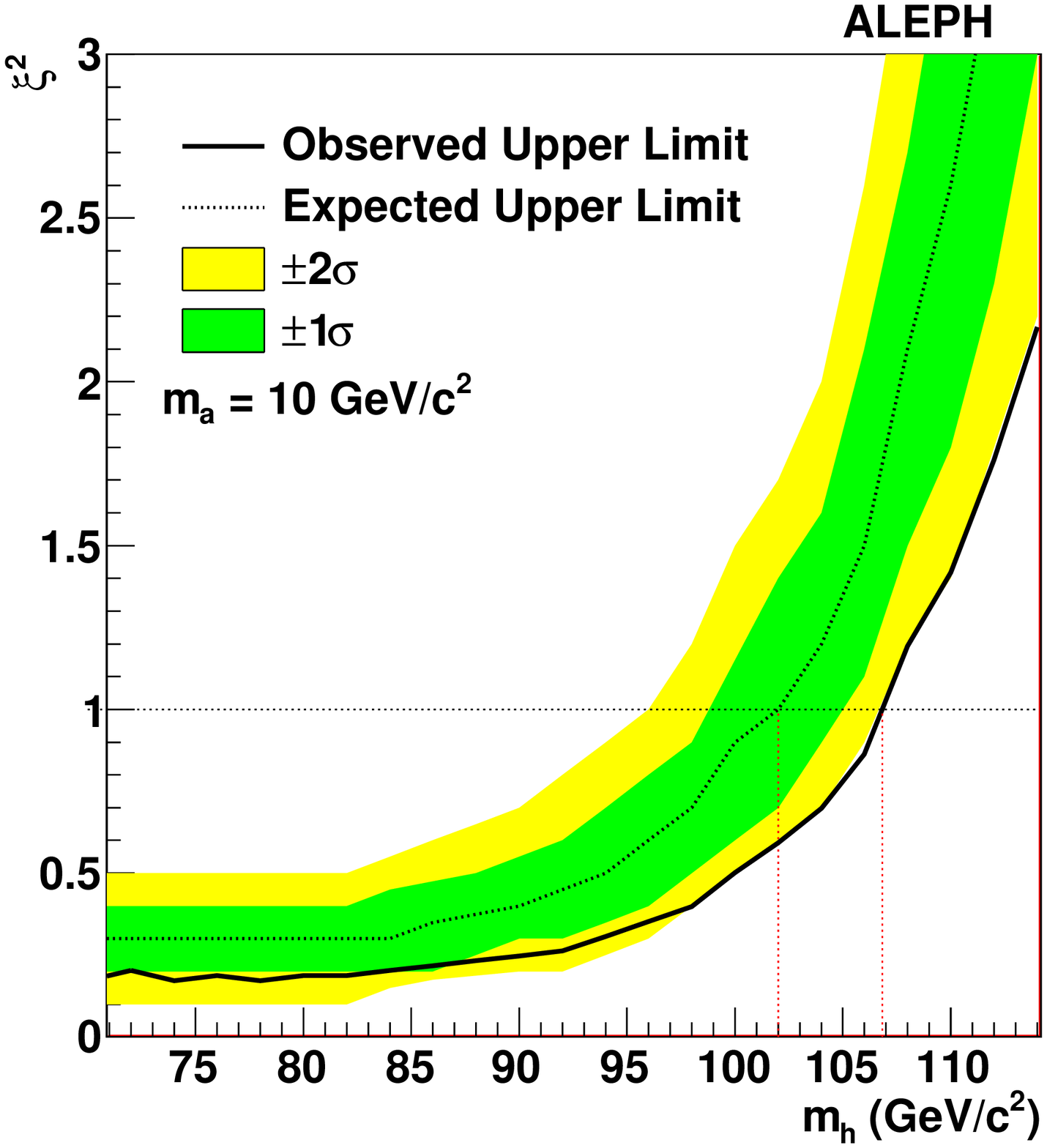}
\end{minipage} 
\begin{minipage}{38pc}
\caption{\label{fig:past}  Examples of recently published analyses using older HEP data.
Left: Meaurements of the strong coupling, $\alpha _{s}$ from an event shape analysis of
JADE data at various centre--of--mass energies, $\sqrt{s}$. The full and
dashed lines indicate the result from the JADE NNLO analysis~\cite{jade2}. The results from a
recent NNLO analysis of ALEPH data are also shown~\cite{alephalphas}. Right: Observed and
expected limits from ALEPH on the combined production cross section times branching ratio
in the search for the process $h \rightarrow  2a \rightarrow 4\tau$, as a function of Higgs
boson mass, $m_{h}$~\cite{alephhiggs}.}
\end{minipage} 
\end{figure}

The re-analysis of the JADE experimental $e^{+}e^{-}$ data from the PETRA collider
(DESY, $1979$--$1986$), using a refined theoretical input, state of the art simulation
and new anlaysis techniques has lead to a significant improvement in the determination
of the strong coupling, in an energy range that is still unique~\cite{jade,jade2}.
The running of the strong coupling, in agreement with the QCD prediction
demonstrates the concept of asymptotic freedom~\cite{asym,asym2},
as illustrated in figure~\ref{fig:past}~(left), where the results from a
similar analysis by the ALEPH experiment~\cite{alephalphas} are also shown.


A search for the production and non--standard decay of a Higgs boson in the
LEP collider data (CERN, $1989$--$2000$) was recently published by the ALEPH
Collaboration~\cite{alephhiggs}, where a possible four tau final state is investigated,
resulting from the decays of two intermediate pseudoscalars produced via a
next--to--minimal supersymmetric Standard Model (NMSSM) Higgs
decay~\cite{nmssm,nmssm2}.
For such a model, and for a pseudoscalar mass $m_{a} = 10$~GeV, Higgs
masses $m_{h} < 107$~GeV are excluded at $95\%$ confidence level, as
illustrated in figure~\ref{fig:past}~(right).

\vspace{0.2cm}

\begin{figure}[t]
\begin{minipage}{21.5pc}
\includegraphics[width=21.5pc]{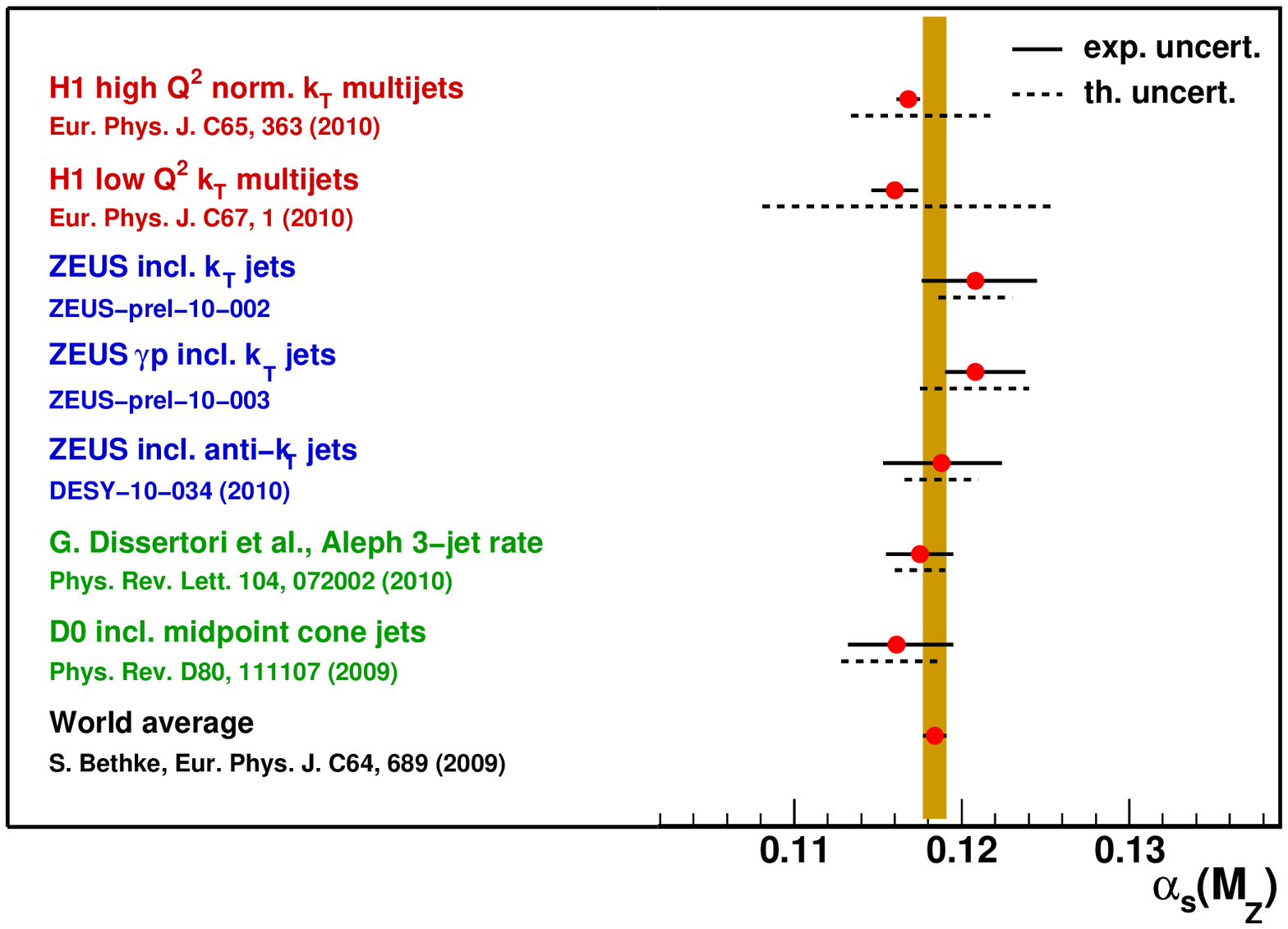}
\end{minipage}
\hspace{1pc}
\begin{minipage}{15.5pc}
\hspace{-0.5cm}
\includegraphics[width=15.5pc]{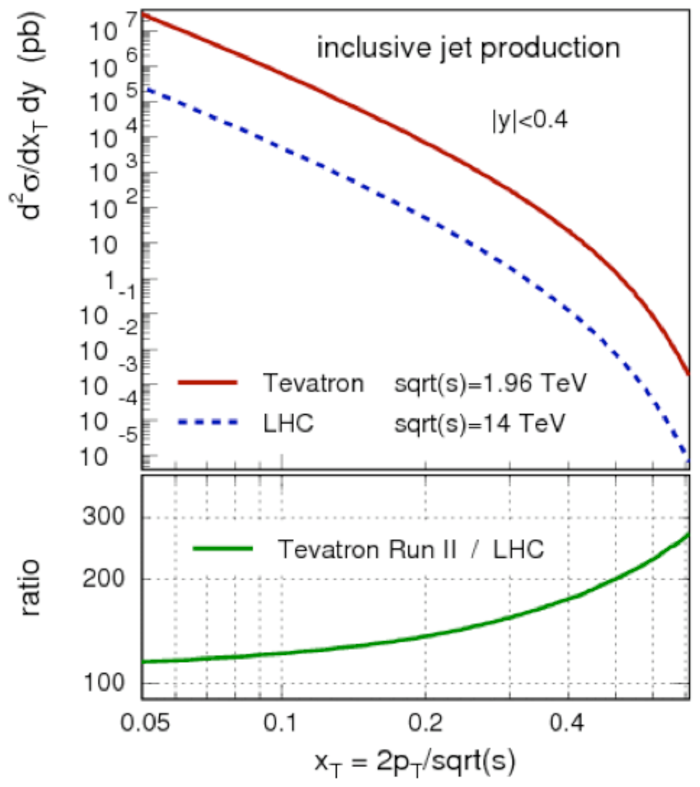}
\end{minipage} 
\begin{minipage}{38pc}
\caption{ \label{fig:physcase} Examples of analyses of current HEP data with potential future impact.
Left: Recent determinations of the strong coupling $\alpha_{s}(M_{Z})$ from a variety of experiments
compared to the $2009$ world average~\cite{kogler}. Right: A comparison of the predicted inclusive jet
production cross sections at the Tevatron and the LHC, as a function of $x_{T}$~\cite{wobisch}.}
\vspace{-0.2cm}
\end{minipage} 
\end{figure}

As discussed in section~\ref{sec:intro}, the $e^{\pm}p$ data from the HERA collider
are themselves a unique achievement, and in many analyses the dominant error on the
measurement is due to the current theoretical uncertainties.
Figure~\ref{fig:physcase}~(left) shows a variety of $\alpha_{s}(M_{Z})$ measuements,
as well as the current world average, where it can be seen that for the latest H1
measurements the theoretical uncertainty domainates the error.
In a situation that mirrors the above JADE analysis, it is hoped that at some point
in the future a better theoretical prediction, including higher order corrections, will be
available inviting the re--analysis of the accurate HERA data.
A similar situation arose recently with the extraction of the strong coupling using
event shape variables by the OPAL Collaboration, where higher order calculations
triggered an improved analysis~\cite{opal}.


The majority of the hadron--hadron particle physics performed at the Tevatron will
eventually be taken over by the LHC, as the amount of $pp$ collision data
at a higher centre--of--mass energy increases.
However, the $p\bar{p}$ collision data taken by the Tevatron experiments will still
be more sensitive to the gluon parton density function (PDF) at high Bjorken $x$ for
some time, where the production cross section for central jets at high
$x \propto x_{T} = 2 P_{T} / \sqrt{s}$ is substantially larger at the Tevatron compared
to at the LHC~\cite{nunnemann}.
A comparison of inclusive jet production cross section predictions from the Tevatron
and the LHC is shown in figure~\ref{fig:physcase}~(right)~\cite{wobisch}.

Another assumption is that the physics potential is exhausted at the end of
the experimental programme.
However, the available person power usually decreases rapidly towards the
end of an experiment, which results in $5$--$10\%$ of the publications 
being finalised at a later stage, when an archival mode of analysis is performed.
This scenario is< true of the LEP papers, where the publication timeline exhibits
a long tail extending well beyond the end of data taking~\cite{travistail}.
Indeed, the above mentioned Higgs analysis is part of this tail.
Interestingly, the predicted publication timescale for the remaining BaBar analyses
also shows the same feature~\cite{babar}.


Drawing on these examples, several scenarios exist where the preservation of
experimental HEP data would be of benefit to the particle physics community:
An extension of the existing physics programme may be necessary to ensure the
long term completion of ongoing analysis;
It may be favourable to re-do previous measurements to achieve an increased precision:
reduced systematic errors may be possible via new and improved theoretical calculations
(MC models) or newly developed analysis techniques;
Preserving old data sets may allow the possiblility to make new measurements at energies
and processes where no other data are available (or will become available in the future);
Finally, if new phenomena are found in new data at the LHC or some other future collider, it may
be useful or even mandatory to go back, if possible, and verify such results using older data.

\section{Models of Data Preservation}
\label{sec:models}

The resurrection of the JADE anlaysis chain to perform the analyses
described above, carried out in the late $1990$'s many years after the
end of data taking, proved to be an eventful exercise and often a subject
of luck rather than careful planning~\cite{bethkeqcd10}.
The general status of the LEP data, which was recorded as recently as the year
$2000$, is a concern, despite the continued paper output.
A recent review of the status of the data of the four experiments identified
that efforts are needed to ensure long term access~\cite{lep}.
The implementation of a data preservation model as early as possible
in the lifetime of an experiment may greatly increase the likelihood of
success, minimise the effort and ease the use of the data in the final
years of the collaboration.


In order to identify different models of data preservation, first an important
question must be asked: What is HEP data?
The data themselves, the digital information in the event files and in databases,
are only a small part of the complete picture: data preservation is not just
about the data! 
Indeed, discussions within the DPHEP study group suggest for pre--LHC experiments
a total of between on half and a few PB of data should be preserved, such that today's
computing centres are, at least by volume arguments, able to store the
data\footnote{The collisions recorded by the LHC experiments result in $10$'s of
TB of data per day, or up to $15$~PB per year.}.
In addition, the various software (simulation, reconstruction, analysis, user)
must be considered.
Concerning documentation, publications of data analysis or detector studies
may be in journals, on SPIRES or arXiv, in HEPDATA or some other database,
and may take the form of full papers, notes, manuals or slides.
Many types of internal meta-data may also exist.
The unique expertise of collaboration members is also at risk, as the person power
associated to the experiment decreases.
By planning a transition of the collaboration structure to something more
suited to an archival mode, this particular loss may be minimised (see section
\ref{sec:governance}).


The different data preservation models established by DPHEP are summarised
in table~\ref{tab:models}, organised in levels of increasing benefit, which comes
with increasing complexity and cost.
Each level is associated with use cases, and the preservation model adopted
by an experiment should reflect the level of analysis expected to be available
in the future.
More details on each of the preservation levels is given in the first DPHEP
publication~\cite{dpheppub1}.

\begin{table}[h]
\begin{center}
 \renewcommand{\arraystretch}{1.2} 
\begin{tabular}{|p{7.2cm}|p{7.2cm}|}
\hline
Preservation Model  & Use Case \\
\hline                                        
\hline
1.~Provide additional documentation &  Publication-related information search\\
\hline
2.~Preserve the data in a simplified format & Outreach, simple training analyses\\
\hline
3.~Preserve the analysis level software and the data format & Full scientific analysis based on existing reconstruction\\
\hline
4.~Preserve the reconstruction and simulation software and basic level data & Full potential of the experimental data\\
\hline
\end{tabular}
\end{center}
\vspace{-0.35cm}
\caption{\label{tab:models} Various data preservation models, listed in order of increasing complexity.
Subsequent models are inclusive: e.g. model 4 also includes the steps and use cases of models 1,2 and 3.}
\end{table}


Past experiences with old HEP data like those described in section~\ref{sec:physcase}
indicate that the definition of the data should include all the necessary
ingredients to retrieve and understand it in order to perform new analyses and that a complete
re--analysis is only possible when all the ingredients can be accounted for.
Only with the full flexibility does the full potential of the data remain, equivalent
to the DPHEP level $4$ data preservation.
Accordingly, the majority of participating experiments in the study group plan
for a level $4$ preservation programme, although different approaches are
employed concerning how this goal can be achieved.


Although a level $1$ preservation model, to provide additonal documentation, is considered
the simplest, this still requires some, often substantial, activity by the experiment.
The HERA collaborations, as well as BaBar, are all currently involved in dedicated efforts
to safeguard and streamline the available documentation concerning their respective
experiments.
A level $2$ preservation, to the conserve the experimental data in simplfied format,
is considered to be unsuitable for high level analysis, lacking the depth to allow, for
example, detailed systematic studies to be performed.
However, such a format is ideal of education and outreach purposes, which many
experiments in the study group are also actively interested in (see section \ref{sec:outreach}).

\section{Common Data Preservation Projects}
\label{sec:projects}

Since the formation of  DPHEP, and especially after the initial findings
of the group were published, the activities and models of the experiments have
aligned to a certain degree and joint initiatives have been launched, related to all
four data preservation levels.
These projects are described in the following.

\subsection{A generic validation suite}
\label{sec:validation}

For data preservation to be truely useful, not only the data themselves must be preserved, 
but also the ability to perform some kind of meaningful operation on them.
In the case of HEP, this means preserving the software and environment employed to
analyse the data (level $3$ preservation model), or if the reconstruction
software is also included, a model where the data or Monte Carlo maybe reproduced
(level $4$ preservation model).
While freezing the software in the current state is an option, experience has shown that
this strategy would sustain analysis capability for only a limited amount of time,
as well as introducing limitations by design.
In order to preserve analysis capabilities for longer periods it would be beneficial to
migrate to the latest software versions and technologies for as long as possible.
Given the pace of technological changes, concerning multi--core CPU design,
changing storage models and system architechtures, as well the dependence on
infrastructures such as the GRID or Clouds, and their associated protocols,
this is a challenging prospect\cite{bogdan}.

\begin{figure}[t]
\begin{center}
\includegraphics[height=36pc, angle=90]{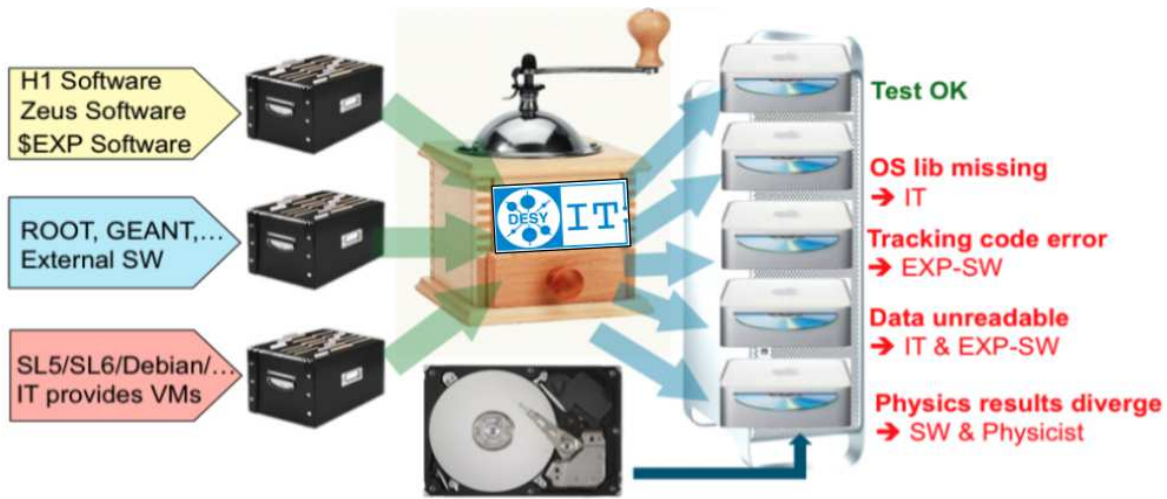}
\end{center}
\begin{minipage}{38pc}
\caption{ \label{fig:validation} A sketch of the proposed experimental software validation scheme at DESY-IT.}
\vspace{-0.4cm}
\end{minipage} 
\end{figure}

It would therefore be beneficial to have a framework to automatically test 
and validate the software and data of an experiment against changes and
upgrades to the environment, as well as changes to the experimental software.
As such a framework would examine many facets common to several current
HEP experiments interested in a more complete data preservation model,
the development of a generic validation suite is favourable.
A test version of such a suite, which includes automated software build tools
and data validation, is currently implemented at DESY-IT, in co-ooperation with
the HERA experiments\cite{janusz}.
The scheme, which is illustrated in figure~\ref{fig:validation} is realised using
a virtual environment capable of hosting an arbitrary number of virtual machine
images, where the inputs to the images are separated into three catagories:
experimental software, external software and operating system.
An image is built with different configurations of operating systems and the relevant
software, and pre-defined tests are then performed to detect migration problems and
incoherence,  as well as identifying the reasons behind them.
Such a framework is by design expandable and able to host and validate the requirements
from multiple experiments.
A full version of the validation suite may now be implemented at DESY-IT,
to safeguard the HERA data for the long term.

\subsection{Global documentation initiatives}
\label{sec:inspire}

As well as the afforementioned individual documentation efforts, global information
infrastructures in HEP may be beneficial to the data preservation project.
INSPIRE~\cite{inspirenet}, the successor to SPIRES, is an existing third-party 
information system for HEP, and is thus ideally situated to provide external
management of experimental documentation.
As well as many overall improvements~\cite{inspire1} with respect to the ageing
SPIRES system, the INSPIRE project is preparing for the ingestion of much more
high--level information in addition to the scientific papers themselves.
These additions range from simple, documented information from the experiments
about a given analysis, through wikis and news-forums, to even the data themselves,
in a storage model where controlled access is possible.

\begin{figure}[t]
\begin{center}
\includegraphics[height=36pc, angle=90]{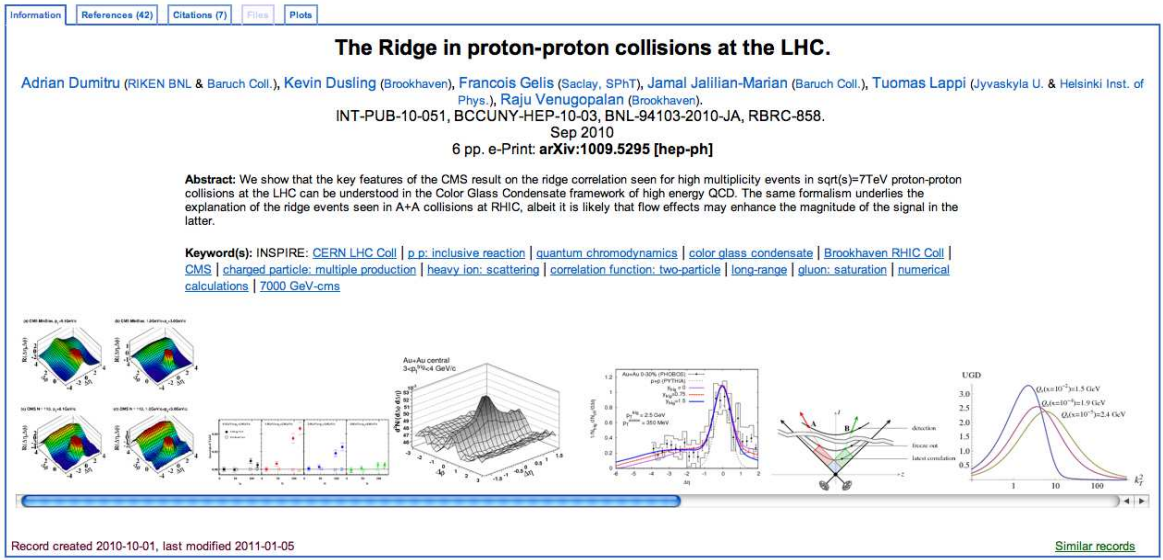}
\end{center}
\begin{minipage}{38pc}
\caption{ \label{fig:inspire} An example of the new record layout for papers in the INSPIRE database~\cite{inspire2}.
An additional tab is foreseen for internal information, visible only to the author and/or collaboration.}
\vspace{-0.4cm}
\end{minipage} 
\end{figure}

In a collaboration with the H1 experiment, INSPIRE will trial a few projects such as the
hosting of H1 internal notes, and the linking of paper histories to publication records.
This idea enables the presentation of the full history of a scientific result,
from the initial conference presentations and papers, through internal talks and notes, 
to a final submitted and refereed publication.
Another major advantage of such a scheme is that the responsibility of hosting the
information passes from a defunct experiment to an active environment.
An example of a new INSPIRE publication record~\cite{inspire2} is shown in
figure~\ref{fig:inspire}, where additional information would appear as an extra tab,
which may, if desired, be only visible to collaboration members.
There are clearly many possibilities for the experiments and INSPIRE to work
together, and more fruitful collaborations are expected via the DPHEP study group.

\subsection{HEP data for outreach, education and open access}
\label{sec:outreach}

The development of a HEP data format for outreach and education,
equivalent to the DPHEP level $2$ data preservation model, is an attractive proposition.
In most cases such a project would run in parallel to preserving the full
re--analysis potential.
In recent years there is a notably increased global effort to improve the
overall level of education in particle physics and to provide access to HEP
to more people than ever before.
Websites such as {\it Teilchenwelt}~\cite{teilchenwelt} or {\it Quarknet}~\cite{quarknet},
as well as the {\it LHC@home} project~\cite{lhcathome}, help further the public 
understanding of science.
Tutorials using a simplified format of real HEP data would be the next
logical step, presented as HEP data with associated pedagogical exercises.
Such a scheme has started within the BaBar Collaboration~\cite{matt} and
following recent discussions within DPHEP about common data formats,
a true, global HEP data portal for outreach purposes seems possible.
The Belle Collaboration also have an outreach programme, {\it B--Lab}, aimed at
high school students, which uses real experimental data~\cite{blab}.


The challenge of releasing such formats to the public domain is to provide
useful open access of HEP data beyond the walls of the original collaboration.
There are however, many issues to consider, such as control of the data, correctness
and reputation of the experiment, not to mention a lack of portability and
the typical state of the documentation within the collaboration.
The implications of open access need to be considered by the collaboration and the
importance of a coherent strategy and presentation of the HEP data when
it is published must be emphasised.

\section{Resources, Governance and Structures}
\label{sec:governance}

The transition to a data preservation model should be planned in advance of the
projected end date of an experiment.
An early preparation is needed and sufficient resources should be provided in
order to maintain the capability to re-investigate such older data samples.
However, the additional resources are estimated to be rather small in comparison
to the person power allocated during the running period of an experiment.
Typically, a surge of $2$--$3$ FTEs for $2$--$3$ years, followed by steady $0.5$--$1.0$
FTE per year per experiment is required for the implementation of a level $3$ or $4$ preservation
model, which should be compared to $300$--$500$ FTEs per experiment for many years.
Therefore, the data preservation cost estimates represent typically much less than $1\%$ of
the original investment, for a potential $5$--$10\%$ increase in physics output.
 

The future structure of a collaboration should also be considered by HEP experiments.
If the transition to a long term analysis model is begun too late the experimental
organisation also risks being left in an undefined state.
In particular, the scientific supervision of the preserved data sets and decisions regarding
authorship and access to data, affecting potential outreach projects, may benefit
from a restructuring of the collaboration towards the final years.
The presence and influence of DPHEP may facilitate this transition, as an interface
to global HEP solutions and as an aid to form common policy and standards.

\begin{figure}[t]
\begin{center}
\includegraphics[height=29pc, angle=90]{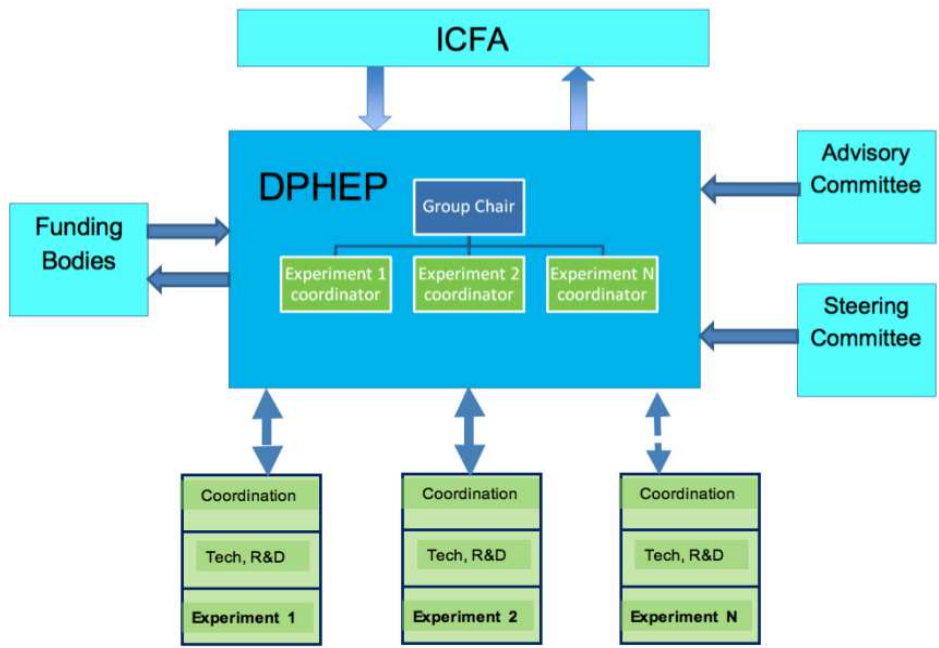}
\end{center}
\begin{minipage}{38pc}
\caption{ \label{fig:dphepstructure} A sketch describing the structure and interactions of the DPHEP Study Group.}
\vspace{-0.4cm}
\end{minipage} 
\end{figure}

Support for the DPHEP initiative has been expressed by CERN, DESY, Fermilab,
IHEP and SLAC, as well as a variety of HEP committees.
The lightweight structure of DPHEP and its interfaces is illustrated in
figure~\ref{fig:dphepstructure}.
Representatives from the laboratories, the experiments and the computing centres,
who are officially appointed by their organisations, are present, with one individual
appointed by  ICFA to take chairmanship of the group.
The organisation receives input from an advisory board, representing all
stakeholders, and continues to ultimately report to ICFA. 
The consolidation and continuation of the international cooperation within DPHEP
is essential to the success and viability of the the data preservation effort.

\section{Conclusions and Outlook}
\label{sec:outlook}

The collection of high energy physics data represents a significant investment
and physics cases can be made to demonstate the potential for scientific results
beyond the lifetime of a collaboration.
However, until recently no coherent strategy existed regarding long term access of HEP data and
an international study group, DPHEP, was formed to address this issue in a systematic way.
Given the current experimental situation, data preservation efforts in HEP are timely, and
large laboratories should define and install data preservation projects in order to avoid
catastrophic loss of data once major collaborations come to an end.
The preservation of the full analysis capability of experiments, including the reconstruction
and simulation software, is recommended in order to achieve a flexible and meaningful
preservation model.
Such a project requires a strategy and well--identified resources, but provides additional research
at particularly low cost, enhancing the return on the initial investment in the experimental facilities.


The efforts of the group are best performed within the common organisation at the international
level DPHEP, through which there is a unique opportunity to build a coherent structure for the future.
Common requirements on data preservation are now evolving via DPHEP into inter--experimental
R\&D projects, optimising the development effort and potentially improving the  degree of
standardisation in HEP computing in the longer term.
The next DPHEP workshop is at Fermilab in May 2011 and a second publication is
expected shortly from the group, describing the current projects in more detail
and providing recommendations and guidelines for future HEP experiments.


\end{document}